\documentclass[conference]{IEEEtran}

\usepackage{blindtext, graphicx}
\usepackage[backend=bibtex,style=ieee,sorting=ynt]{biblatex}
\usepackage[T1]{fontenc}

\ifCLASSOPTIONcompsoc
  \usepackage[caption=false,font=normalsize,labelfont=sf,textfont=sf]{subfig}
\else
  \usepackage[caption=false,font=footnotesize]{subfig}
\fi

\usepackage[cmintegrals]{newtxmath}

\usepackage{float}
\usepackage{amsmath}
\usepackage{varwidth}
\usepackage{float}

\usepackage[ruled,vlined,linesnumbered]{algorithm2e}
\usepackage{amsmath}
\usepackage[noend]{algpseudocode}
\IEEEoverridecommandlockouts

\makeatletter
\def\BState{\State\hskip-\ALG@thistlm}
\makeatother

\usepackage{tikz}   
\usepackage{siunitx} 
\usepackage{pgfplots}
    \usetikzlibrary{
        pgfplots.colorbrewer,
    }
    \pgfplotsset{
        cycle list/.define={my marks}{
            every mark/.append 
            style={solid,fill=\pgfkeysvalueof{/pgfplots/mark list fill}}\\
            every mark/.append style={solid,fill=\pgfkeysvalueof{/pgfplots/mark list fill}},mark=square*\\
            every mark/.append style={solid,fill=\pgfkeysvalueof{/pgfplots/mark list fill}},mark=triangle*\\
            every mark/.append style={solid,fill=\pgfkeysvalueof{/pgfplots/mark list fill}},mark=diamond*\\
            every mark/.append
            style={solid,fill=\pgfkeysvalueof{/pgfplots/mark list fill}},mark=*\\
        },
    }

\begin{document}

\title{Physical Layer Communications System Design Over-the-Air Using Adversarial Networks}

\author{\IEEEauthorblockN{Timothy J. O'Shea}
\IEEEauthorblockA{DeepSig Inc \& Virginia Tech\\
Arlington, VA\\
toshea@deepsig.io}
\and
\IEEEauthorblockN{Tamoghna Roy}
\IEEEauthorblockA{DeepSig Inc.,\\
Arlington, VA\\
troy@deepsig.io}
\and
\IEEEauthorblockN{Nathan West}
\IEEEauthorblockA{DeepSig Inc.,\\
Arlington, VA\\
nwest@deepsig.io}
\and
\IEEEauthorblockN{Benjamin C. Hilburn}
\IEEEauthorblockA{DeepSig Inc.,\\
Arlington, VA\\
bhilburn@deepsig.io}
}

\maketitle
\begin{abstract}
This paper presents a novel method for synthesizing new physical layer modulation and coding schemes for communications systems using a learning-based approach which does not require an analytic model of the impairments in the channel.   It extends prior work published on the \textit{channel autoencoder} to consider the case where the channel response is not known or can not be easily modeled in a closed form analytic expression.  By adopting an adversarial approach for channel response approximation and information encoding, we can jointly learn a good solution to both tasks over a wide range of channel environments.  We describe the operation of the proposed adversarial system, share results for its training and validation over-the-air, and discuss implications and future work in the area.
\end{abstract}


\IEEEpeerreviewmaketitle

\section{Introduction}

\begin{figure}
    \centering
    \includegraphics[width=0.5\textwidth]{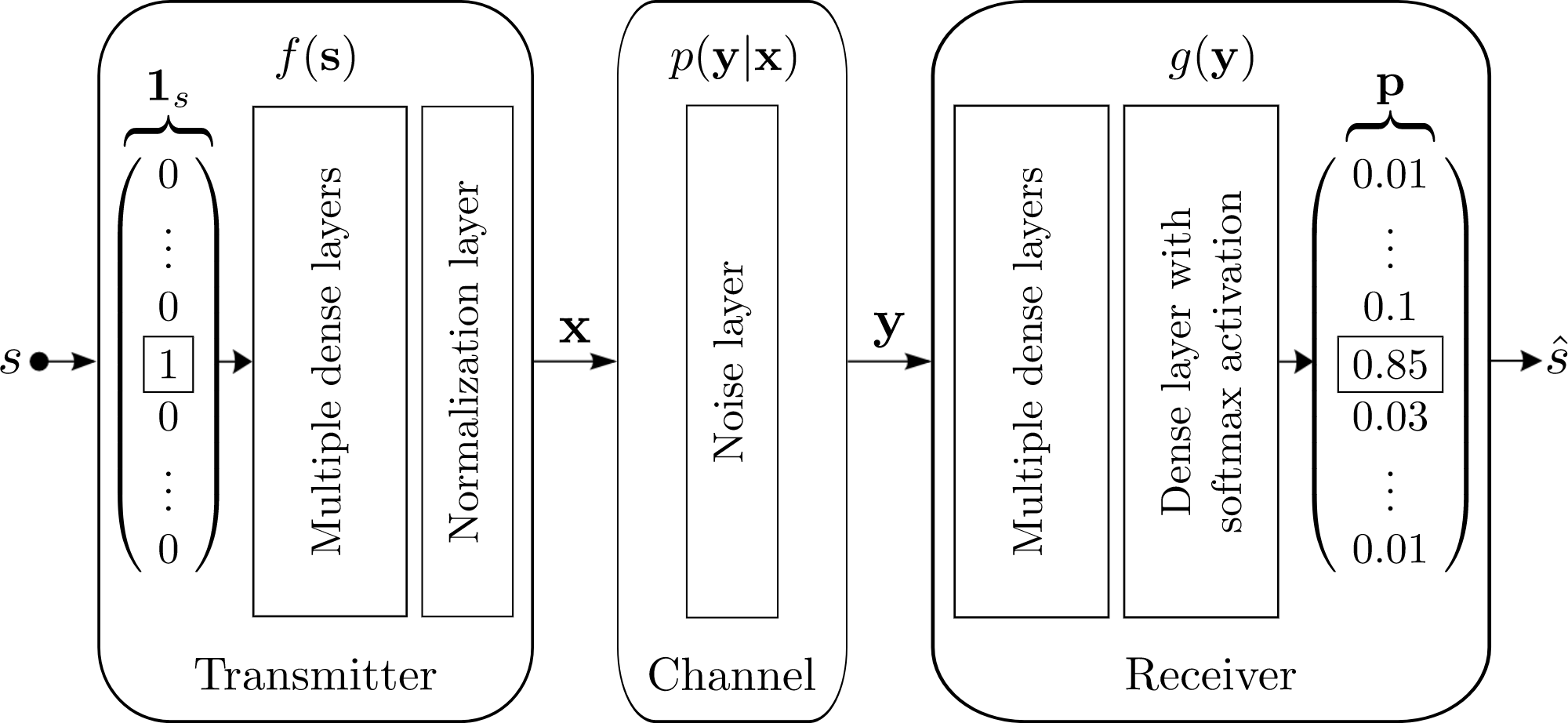}
    \caption{A channel autoencoder system for learning physical layer encoding schemes optimized for a differentiable analytic channel model expression}
    \label{fig:cae}
\end{figure}

The channel autoencoder \cite{o2017introduction,o2016learning} (shown in Figure \ref{fig:cae}) has recently received significant interest as a method for designing communications systems' physical layer information encoding schemes.  By jointly optimizing large multi-layer non-linear encoder and decoder networks with many degrees of expressive freedom, made possible by techniques collectively known as deep learning \cite{goodfellow2016deep}, it has been shown that effective new modulation schemes with an inherent error-correction capability can be readily learned for a variety of common channel models.

This is an exciting result in that rather than manually designing radio modulation schemes for analytic representation convenience (e.g., slicing on nice rectangular grid boundaries) and then measuring or demonstrating their optimal use, we can directly learn modulation and demodulation functions of high complexity which optimize the complete system for global performance metrics or loss functions (e.g., symbol error rate).  This method can provide highly efficient solutions for analytic channel models of single-user, multi-user, and multi-antenna communications channels.  To optimize for the effects of a real-world system, however - including the effects induced by digital conversion, analog RF hardware, and other sources of distortions and impairments - the analytic expression used within the channel model must accurately represent all of these effects, and it must do so in a differentiable way suitable for backpropagation.  

Unfortunately, for many systems, it is difficult to accurately capture all of these effects in a closed-form analytic representation.  Thus, they are often represented using simplified models which cannot fully express real-world complexities (e.g., device-to-device hardware variation).  For this reason, it is highly desirable to learn a radio communication scheme directly from the sampled response of real physical hardware devices and channels rather than attempting to model the responses manually.  One solution to this, which is implemented and explored in \cite{dorner2017deep}, is to perform gradient updates only on the receiver network after synthetic pre-training. However, due to the lack of a channel gradient expression, this approach provides no way to further update the transmitter or modulation scheme.  Differentiation through the wireless void is difficult. An approach using reinforcement learning and sampling is proposed in \cite{grathwohl2017backpropagation} for related problems. Our approach, as presented in this paper, instead treats the problem as an adversarial function approximation problem.

In the next section we describe our proposed method for fully adapting the transmitter and receiver to a physical model-free channel response using a hardware-in-the-loop generative adversarial channel autoencoder network over-the-air.

\section{Technical Approach}

\begin{figure*}
    \vspace{0.2in}
    \centering
    \includegraphics[width=0.80\textwidth]{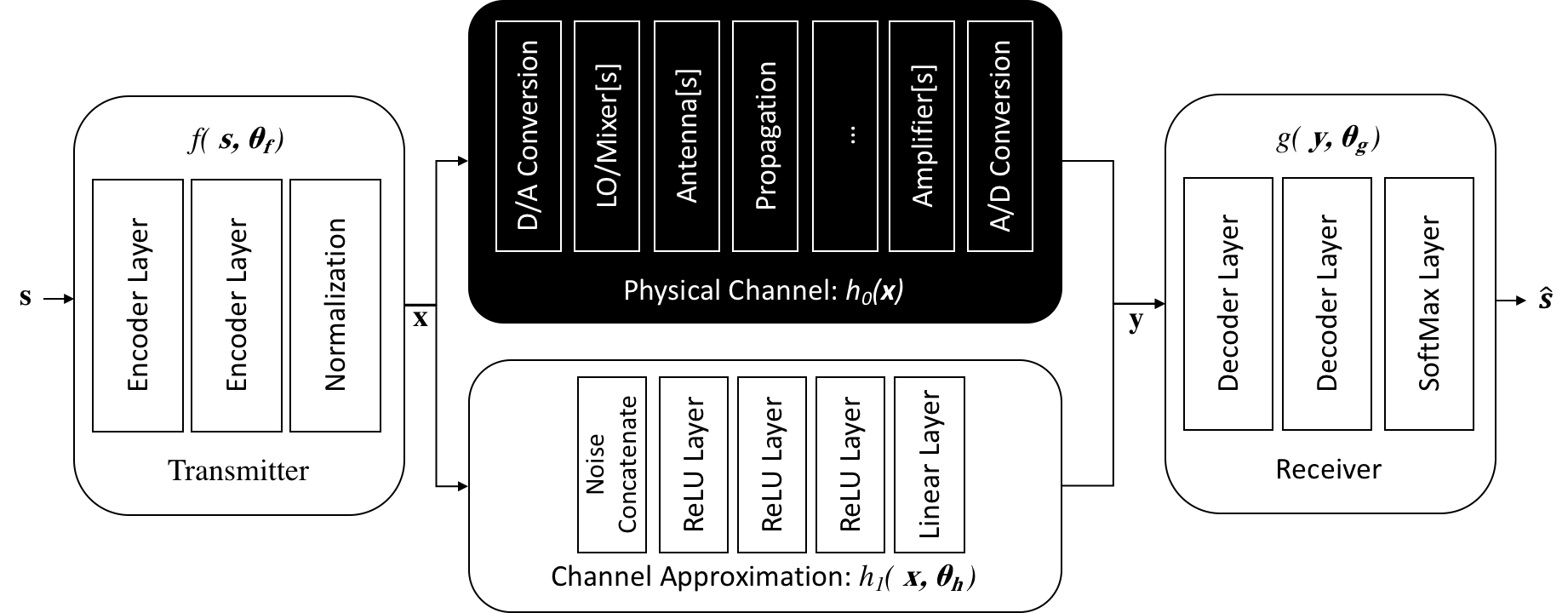}
    \caption{A generative adversarial network for learning a communications system over a physical channel with no closed-form model or expression}
    \label{fig:acae}
\end{figure*}

Generative adversarial networks (GANs) \cite{goodfellow2014generative} have recently been used effectively for a number of applications, including generating false images which confuse learned image recognition systems into mis-classifying an object.  They work by simultaneously training multiple adversarial objective functions, simultaneously or iteratively, which inform and improve each others' performance for some task. In the false image generation case, the basic approach trains a discriminator whose objective is to minimize the error in classifying images as \textit{real} or \textit{fake}, and a second network whose objective is to minimize the error in generating images from noise which classify as \textit{real}.  Since the original work in \cite{goodfellow2014generative}, numerous variations of this approach have significantly improved the performance of GANs \cite{ganzoo}, but they generally use the same original concept.  Building upon this work, we combine our prior work on channel autoencoders with ideas inspired by GAN research to jointly optimize for the two tasks of:

\begin{enumerate}
    \item approximating the response of the channel in any arbitrary communications system, and
    \item learning an optimal encoding and decoding scheme which optimizes for some performance metric (e.g., low symbol error rate)
\end{enumerate}

The basic configuration of this approach, which we refer to as a Communications GAN (CommGAN), is shown in Figure \ref{fig:acae}.   Here, as with the original channel autoencoder shown in Figure \ref{fig:cae}, we leverage two networks for encoding and decoding of information symbols ($s$) which comprise our communications transmitter and receiver.  The encoder network, $f(s,\theta_f)$, encodes codeword indices or bits into transmit waveform (sample values) or constellation (I/Q) values using a set of encoder weights $\theta_f$ through a series of neural network layers. The encoder network comprises fully connected layers with a ReLU activation, which enables non-linear transformations.  The decoder network, $g(y,\theta_g)$, does the inverse - mapping the set of received samples (i.e., digitized voltage levels at the receive antenna) into a set of likelihood levels for each bit or codeword which may have been transmitted. The decoder network also comprises a series of fully connected layers, again using ReLU for non-linear transforms, and a SoftMax output activation.  In contrast to our original channel autoencoder work, we do not employ a channel model such as an analytic expression for Additive White Gaussian Noise (AWGN) or Rayleigh fading, but instead introduce two forms of the channel $h(x)$ to encompass modeling of any black-box channel transform: 

\begin{enumerate}
    \item $h_0(x)$: A real-world physical measurement of the response of a communications system comprising a transmitter, a receiver, and a channel.
    \item $h_1(x,\theta_h)$: A non-linear multi-layer neural network which seeks to mimic the channel response of $h_0$ synthetically.
\end{enumerate}

Throughout the training process, we use each of these channel expressions to iteratively arrive at an optimized solution.  The simplified training approach used to train each of these networks is described in pseudo-code in Algorithm \ref{training}.  Initially, we use a Mean-Squared Error (MSE) channel approximation loss ($\mathcal{L}_0$), and a sparse categorical cross-entropy channel autoencoder loss ($\mathcal{L}_1$).

\begin{algorithm}
\caption{CommGAN Training Routine\label{training}}
\DontPrintSemicolon
$ \theta_f,\theta_h,\theta_g \leftarrow \text{random initial weights} $ \;
\ForEach{epoch in epochs}{
    \ForEach{$step$ in $steps_0$}{
        $ s_0 \leftarrow \text{random symbol values} $ \;
        $ y_0 =       h_0(f(s_0,\theta_f))           $ \;
        $ \hat{y}_0 = h_1(f(s_0,\theta_f),\theta_h)  $ \;
        $ \theta_h \leftarrow \text{Adam Update}(\mathcal{L}_0(y_0,\hat{y}_0)) $ \;
    }
    \ForEach{$step$ in $steps_1$}{
        $ s_0 \leftarrow \text{random symbol values}   $   \;
        $ \hat{s}_0 = g(h_1(f(s_0,\theta_f)),\theta_g) $ \;
        $ \theta_f,\theta_g \leftarrow \text{Adam Update}(\mathcal{L}_1(s_0,\hat{s}_0)) $ \;
    }
    Display metrics \;
    Check stopping criteria
    }
\end{algorithm}

Here we simply cycle between objectives, updating weights for each network during the appropriate stage with manually tuned learning rates and relatively small networks for $f$, $g$, and $h$, employing several fully connected ReLU layers for each.  

The physical channel $h_0(x)$ is implemented using a Universal Software Radio Peripheral (USRP B210) \cite{ettus2009universal} and custom software tooling. The sample clocks of the transmit and receive hardware share a clock reference, but the devices have uncalibrated offsets and impairments from things like group delay, analog components, antenna and amplifier distortion, and over-the-air effects.  We operate the radios in the 900 MHz ISM band using omni-directional whip antennas in a relatively benign indoor laboratory environment at a sample rate of 1 MSamp/sec.  We loosely calibrate the transmit and receive times based on a rising edge pulse to within about 1 sample (1 microsecond), where fractional timing error and channel effects exist between the transmit and receive samples.  For purposes of the encoder and decoder networks, we map information to 16 discrete symbols (i.e., 4 bits per symbol). We use 3 samples per symbol on the transmitter, and in each stage we send the three samples from one symbol and consider a single receive sample at the receiver, which occurs somewhere within the unaligned 3-sample time window (i.e., there is an un-calibrated symbol offset between the transmitter and receiver).

\section{Results} 

We found that normalization and noise insertion are important aspects of network training, just as they were in the fully simulated channel autoencoder implementation.  Effects such as clipping (i.e., 'saturation') in physical analog-to-digital converters are helpful to capture in normalization to improve training convergence. Including an average power constraint, like those used in simulation, seem to mitigate un-helpful gradients - for instance, solutions which simply continually increase the transmit power.

\begin{figure}
    \centering
    \includegraphics[width=0.47\textwidth]{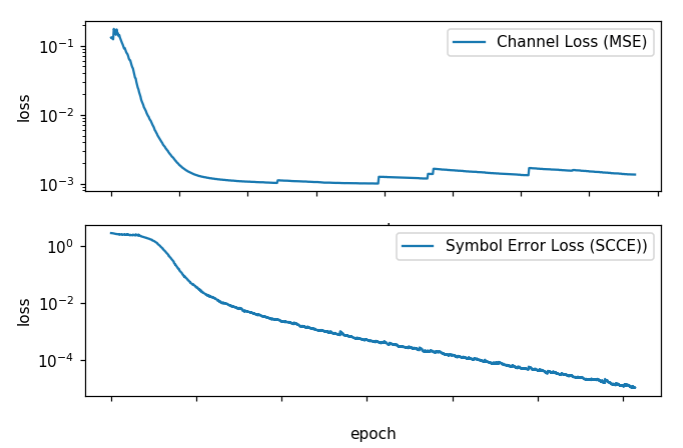}
    \caption{CommGAN loss curves during the training process}
    \label{fig:loss}
\end{figure}

The training process completes in less than five minutes using an NVIDIA Titan V GPU. Figure \ref{fig:loss} presents the loss curves of the channel approximation accuracy and the autoencoder symbol cross-entropy during the training process.  Here, both functions reinforce the other, iteratively building better models of the channel and learning more optimal transmit signal representations as time and the number of epochs increase.

\begin{figure}
    \centering
    \includegraphics[width=0.47\textwidth]{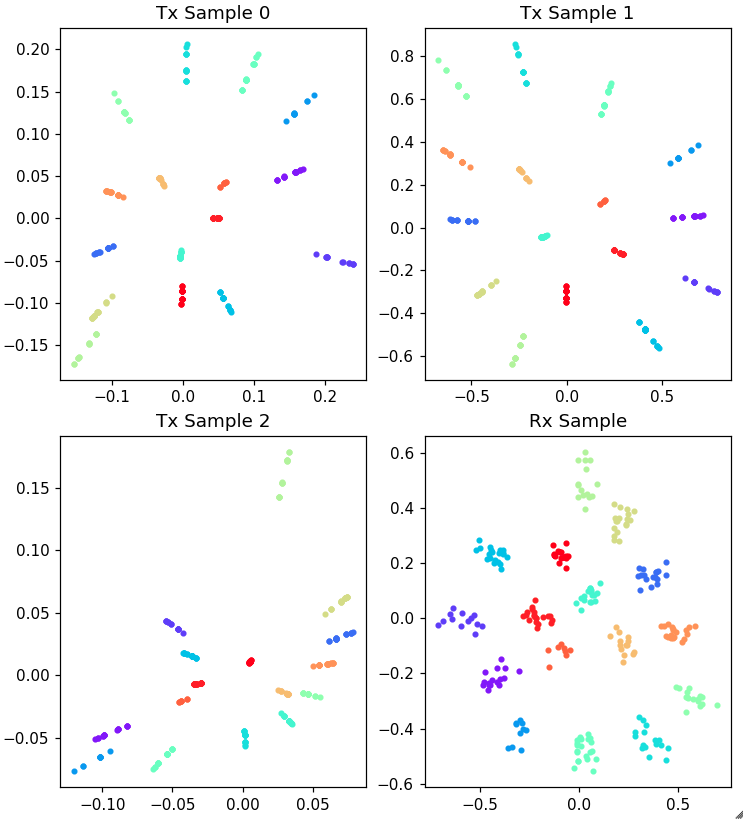}
    \caption{Learned CommGAN scheme over the 900 MHz ISM band using a USRP B210 at 1 MSamp/sec, encoding 4 bits into one receive symbol from 3 unaligned over-the-air transmit symbols}
    \label{fig:const}
\end{figure}

The resulting learned modulation scheme is a non-standard 16-QAM mode which achieves a symbol error rate of around 0.00714, measured over-the-air.  Figure \ref{fig:const} plots the encoding scheme's learned I/Q samples after training, showing the three complex transmitted samples and the single complex received sample, taken at some fractional time within the 3-sample time window, with an over-the-air channel.  We can see that symbols 0, 1, and 2 learn to transmit similar constellations with varying scaling and rotation, which interpolate to form a clean non-standard 16-QAM receive constellation, shown in the bottom right, with a different rotation relative to the transmitted constellations.

\section{Conclusion} \label{conclusion}

We have shown in this paper that by using an adversarial approach we can learn function approximations for arbitrary communications channels, and that by jointly learning a channel function approximation and an encoder/decoder scheme we can learn effective communication systems which achieve robust performance without needing a closed-form channel model or implementation.  Previously, over-the-air channel autoencoders required pre-training based on a closed-form model designed to match the expected deployment scenario and could not back-propagate through the black-box void of the radio channel, and thus only optimized the receiver side of the network \cite{dorner2017deep}.  Using an adversarial approach, we have shown that such a system can be learned directly on an unknown and uncharacterized physical channel, and that the function approximation for this channel is sufficient to back-propagate and adapt both encoder and decoder networks.

Much future work remains for this approach, such as providing a more in depth analysis of the system performance, addressing training on a stream of information rather than a single symbol (also raised in \cite{dorner2017deep}), investigating how such a system would perform under changing environmental conditions with online over-the-air adaptation, and how it might scale to larger symbol block-sizes on the order of those used by modern LTE cellular standards.

Learning directly from complex systems with high degrees of freedom, such as radio hardware and the propagation effects of physical channels, continues to be a challenging problem for modern communications systems with numerous sources of linear and non-linear impairments which can be difficult to accurately capture with simplified analytic models.  This paper presented an alternative approach which addresses the complexities present in physical radio systems using approximation networks learned directly from data and experience.  While much work is needed to mature this approach, we believe that the ability to optimize a system's performance holistically, over many effects and transforms, is a fundamental requirement for the next major leap in the performance of communications systems.

\printbibliography
\end{document}